\documentclass{aa}
\usepackage{epsfig}
\usepackage{graphicx}
\usepackage{rotating,lscape}
\newenvironment{narrow}[2]{%
  \begin{list}{}{%
  \setlength{\topsep}{0pt}%
  \setlength{\leftmargin}{#1}%
  \setlength{\rightmargin}{#2}%
  \setlength{\listparindent}{\parindent}%
  \setlength{\itemindent}{\parindent}%
  \setlength{\parsep}{\parskip}}%
  \item[]}{\end{list}}
\begin{document}
 
\title{NIR surface photometry of a sample of nearby spiral galaxies}
\author{N. Castro-Rodr\'\i guez\inst{1} \and F. Garz\'on\inst{1,2}}
\institute{Instituto de Astrof\'{\i}sica de Canarias, E--38200 La Laguna, Spain
\and
Departamento de Astrof\'{\i}sica, Universidad de La Laguna, Tenerife, Spain
\and
Astronomisches Institut der Universit\"at Basel,CH--4102, Binningen, Switzerland}

\offprints{ncastro@astro.unibas.ch}

\date{ver. July 3 2003 / Received xxxx / Accepted xxxx}


\abstract{
The first results of an observational programme aimed at mapping a sample of
face--on spiral galaxies in the NIR are presented. This paper shows the
surface photometry of the first ten galaxies in the sample. The data were taken in
the broad band $J$ (1.2 $\mu$m) and $K_{s}$ (2.2 $\mu$m) filters. The sources
were selected
mainly according to their size and brightness in order to suit the
characteristics of the CAIN 2D NIR camera on the 1.5 m Carlos
S\'anchez Telescope (Tenerife, Spain). The primary scientific goal is to
provide a comprehensive and uniform database of the main structural and
photometric parameters of the sample members from NIR surface photometry. To this end, elliptical
isophotal fitting was performed on each galaxy image to extract information
about the size and location of its morphological components and provide the
azimuthally averaged radial brightness profile. Analytical functions for each
component's brightness distribution were then used to match that profile,
and their functional parameters obtained from the global fitting. This first
report includes data for NGC 3344, NGC 3686, NGC 3938, NGC 3953, NGC 4254, NGC
4303, NGC 4314, NGC 5248, NGC 6384 and NGC 7479. 
}

\titlerunning{NIR surface photometry of galaxies}

\maketitle


\

\section{Introduction}

Surface photometry of external galaxies in the optical domain  has burgeoned
 in recent decades with  advent of  large format CCD
arrays and the popularity of the 2D optical cameras. Data have been amassed
since then for a wide range of astrophysical studies, from
morphological classification (van der Kruit \& Searle 1981, 1982) to  stellar population
characterization (de Jong 1994, 1996a, 1996b); from star formation research (Kennicutt 1983, 1989)
to the investigation of the ISM in external galaxies (Valentijn 1994). The
situation is much poorer in other wavelength regimes.
 The near infrared (NIR) domain is
particularly interesting in this respect since in this range the flux is largely
dominated by the direct stellar radiation while drastically reducing  the
general extinction. In addition, NIR colours are very sensitive to population
changes in the galaxy (Peletier et al. 1994; de Jong \& van der Kruit
1994; Giovanardi \& Hunt 1996; Hunt et al. 1997;
Peletier \& Balcells 1996, 1997).

Reasons for the scarcity of NIR extragalactic databases are to be
sought in the late development of 2D detector arrays with respect to their
optical counterparts and in the greater difficulty in building NIR astronomical
instruments, which have to be cryogenically cooled to achieve good performances
in the thermal infrared, namely from the $K$ band towards longer
wavelengths. Furthermore, observations and data reduction are  far more complicated and time
consuming in the NIR than in the optical. The basic reason is that the signals
have to be extracted from a largely dominant sky emission, coming mostly from
the thermal radiation of the earth atmosphere, which in addition is  much more
time-dependent, with variation timescales ranging from several tens of minutes
in the $J$ band, to one or two minutes in $K$. This means that to properly
subtract the sky emission from the image of an extended object, like a galaxy,
one has to nod between two adjacent sky positions with a frequency adjusted to
the  variation timescale of the sky background. Also, the high value of the sky
flux  necessarily shortens the single on-chip integration time to avoid
saturation and/or to keep the counts within the linear range of the detector. The
combined result of all these effects is that an observing run for an extended
object ends up with hundreds of individual exposures that have to be treated
carefully. It is also worth  mentioning that this procedure of sky substraction
permits the galaxy object to completely cover the detector, since the sky will
be measured in the other nodding position. Thus, the criterion of size in the
sample can be extended to the full limit imposed by the instrument's field of
view.

We have recently started an observational programme aimed at producing a 
database of NIR infrared images of spirals galaxies, with several 
primary objectives in mind:

\begin{itemize}

\item To characterize the presence of the relevant morphological components and
compare, whenever possible, their structural parameters with those obtained in
the optical.

\item To check their morphological classification.

\item To investigate trends in structural parameters with morphological type.

\end{itemize}

In this first paper we will not cover the last two items, the sample does
not contain a wide range of galaxy types.
In subsequent reports we will, try to enlarge the sample and investigate the 
underlying stellar
population in order to delineate the presence of interstellar dust by means of NIR
colours.

The galaxies presented in this paper have been selected primarily according  to
their isophotal $D_{25}$ diameter, as listed in the RC3 (de Vaucouleurs 1991),
and  their brightness, both being considered within the suitable range for
testing the feasibility of the project. We included only spirals in this first
sample since they exhibit a wider range of morphological structures.


\
\section{Observations}

The observations were made at the Teide Observatory  (OT) on the island of
Tenerife (Spain) on the 1.5 m Carlos S\'anchez Telescope (CST). The images 
were taken with the common user NIR camera, CAIN, equipped with a NICMOS3
$256^2$ detector array and standard broad band NIR filters $JHK_{\rm s}K$, cooled to
LN$_2$ temperature. CAIN holds two different optical set-ups, 
both fully cryogenic,
which provide two plate scales: the narrow camera, giving 0.4 arcsec per pixel,
and the wide one, with 1 arcsec per pixel, used for this work. All the
galaxies have been observed in at least  two filters, $J$ and $K_{\rm s}$ (or 
$K$), during
several campaigns from 1998 onwards. Typical seeing value was, on average, a little in excess of 1 arcsec.

\begin{table*}[!]
\begin{center}
\begin {tabular}{ccccccccccc}\hline

Galaxy & RA & DEC &  Type &  $T$ &$\log D_{25}$ & PA & $M_B$ & Incl. & $D$ \\ 
       & [h] & [deg] &       &      &   &[$0.1'$] & [deg] & mag & [Mpc] \\ \hline	    
								   
NGC 3344 & 10.72 & 24.92 & SABbc& 4.0& 1.85 &	  & 10.49 &  17.6 & 9.3\\ 
NGC 3686 & 11.46 & 17.22 & SBbc & 4.1& 1.48 & 15  & 11.96 &  40.8 &14.2\\  
NGC 3938 & 11.88 & 44.12 & SAc  & 5.1& 1.69 & 52  & 11.04 &  12.9 &11.4\\ 
NGC 3953 & 11.89 & 52.32 & SBbc & 3.9& 1.88 & 13  & 10.84 &  63.3 &12.6\\  
NGC 4254 & 12.31 & 14.41 & SAc  & 5.2& 1.72 &	  & 10.42 &  32.0 &14.8\\  
NGC 4303 & 12.36 &  4.47 & SABbc& 4.0& 1.79 & 162 & 10.18 &  19.2 &19.4\\  
NGC 4314 & 12.37 & 29.89 & SBa  & 1.0& 1.59 &	  & 11.41 &  19.2 &12.5\\  
NGC 5248 & 13.62 &  8.88 & Sbc  & 4.0& 1.75 & 110 & 11.00 &  50.5 &14.8\\  
NGC 6384 & 17.54 &  7.06 & SABbc& 3.6& 1.69 &  30 & 11.60 &  59.9 &24.1\\  
NGC 7479 & 23.08 & 12.32 & SBc  & 4.4& 1.61 &  25 & 11.77 &  36.4 &34.7\\
\hline
\end {tabular}
\end{center}
\caption{General properties of galaxies taken from the RC3 catalogue. The columns represent: galaxy name; coordinates for equinox 2000;
Hubble type (three columns); diameter, in log units, of the isophotal level of
25 mag per square arcsec; position angle; magnitude in $B$;  inclination angle;
and distance (from Huchtmeier \& Richter 1989).}

\label{Tab:rc3}

\end{table*}

\begin {table*}[!]
\begin{center}
\begin {tabular}{ccccc|ccccc}\hline

Galaxy & Filter & T.exp/frame & Fr./point. & Tot.  point. & 
Galaxy & Filter & T.exp/frame & Fr./point. & Tot.  point. \\ 
       &        & [s] &         &           & 
       &        & [s] &         &           \\ \hline

NGC 3344 & $J$              & 30  & 4   & 21    &  NGC 4303 & $J$              & 20  & 6   & 20  \\   
NGC 3344 & $K_{\rm s}$ & 12  & 10 & 20    &  NGC 4303 & $K_{\rm s}$ & 8    & 15 & 20  \\   
NGC 3686 & $J$              & 20  & 6   & 20    &  NGC 4314 & $J$              & 20  & 6   & 20  \\   
NGC 3686 & $K_{\rm s}$ & 8    & 15 & 20    &  NGC 4314 & $K_{\rm s}$ & 8    & 15 & 20    \\   
NGC 3938 & $J$              & 30  & 4   & 16    &  NGC 5248 & $J$              & 20  & 6   & 20   \\   
NGC 3938 & $K_{\rm s}$ & 10  & 12 & 17    &  NGC 5248 & $K_{\rm s}$ & 6    & 20 & 20  \\   
NGC 3953 & $J$              & 30  & 4   & 20    &  NGC 6384 & $J$              & 30  & 6   & 10    \\   
NGC 3953 & $K_{\rm s}$ & 10  & 12 & 20    &  NGC 6384 & $K$             & 8    & 12 &  10   \\   
NGC 4254 & $J$              & 20  & 6   & 20    &  NGC 7479 & $J$              & 30  & 6   & 16    \\   
NGC 4254 & $K_{\rm s}$ & 8    & 15 & 20    &  NGC 7479 & $K_{\rm s}$ & 12  & 10 & 13      \\

\hline

\end {tabular} 
\end{center}
\caption{Some observing parameters: galaxy name, filter, integration time per
frame, number of frames at each pointing position, number of pointings per
galaxy. The total exposure time on target is the product
of the last three columns on each object.}
\label{Tab:caracteristicas}
\end {table*}

As mentioned in the preceding section, this first sample was selected
taking into account several prescriptions:

\begin {itemize}
\item Declinations between $-35$ and +60 degrees, to be observable from the OT. 

\item Morphological types with T in the range $(-1,9)$. All the galaxies are
spirals (barred and non-barred) and nearly face on (inclination $<60^\circ$) for
easier decomposition of their morphological structures. The majority of the
galaxies in the sample have $T$ in the range  4--5, see table \ref{Tab:rc3}.

\item High surface brightness, $B<13$, and diameters suitable for the telescope
size and the instrument FOV, respectively.

\end {itemize}

The observed sample till now contains more than the ten galaxies, NGC3344,
NGC3686, NGC3938,  NGC3953, NGC4254, NGC4303, NGC4314, NGC5248, NGC6384 and
NGC7479, which constitute the body of this paper
(table \ref{Tab:rc3}). The remaining ones will be presented in
subsequent papers.

\begin {table*}[!]
\begin{center}
\begin {tabular}{cccc|cccc}\hline
Galaxy  &  Filter  &  3$\sigma$  & 5$\sigma$  &
Galaxy  &  Filter  &  3$\sigma$  & 5$\sigma$  \\  \hline
NGC 3344  &  $J$	         &   21.93&  21.37   &  NGC 4303  &  $J$	         &   21.33&  20.78  \\   
NGC 3344  &  $K_{\rm s}$    &   20.34&  19.78   &  NGC 4303  &  $K_{\rm s}$    &   20.17&  19.61   \\  
NGC 3686  &  $J$	         &   22.33&  21.77   &  NGC 4314  &  $J	$                &   22.76&  22.21   \\  
NGC 3686  &  $K_{\rm s}$    &   21.23&  20.68   &  NGC 4314  &  $K_{\rm s}$    &   21.36&  20.80   \\    
NGC 3938  &  $J$	         &   21.59&  21.04   &  NGC 5248  &  $J$	         &   22.08&  21.53   \\  
NGC 3938  &  $K_{\rm s}$    &   20.51&  19.95   &  NGC 5248  &  $K_{\rm s} $   &   21.11&  20.50   \\  
NGC 3953  &  $J$	         &   22.89&  22.33   &  NGC 6384  &  $J$	         &   22.08&  21.53   \\
NGC 3953  &  $K_{\rm s}$    &   21.28&  20.72   &  NGC 6384  & $ K $               &   21.10&  20.56   \\  
NGC 4254  &  $J	$                &   22.52&  21.97  &  NGC 7974  &  $J$	         &   22.54&  21.99   \\  
NGC 4254  &  $K_{\rm s}$    &   19.92&  19.37   &  NGC 7974  & $ K_{\rm s} $   &   20.77&  20.22   \\      
\hline
\end {tabular} 
\end{center}
\caption{Detection limits, in magnitudes per square arcsec, of the final
reduced images in each filter. Values are measured at 3$\sigma$ and 5$\sigma$
 above the background.}
\label{Tab:limite}
\end {table*}		
			
\begin{figure}[!]
\begin{center}
\mbox{\epsfig{file=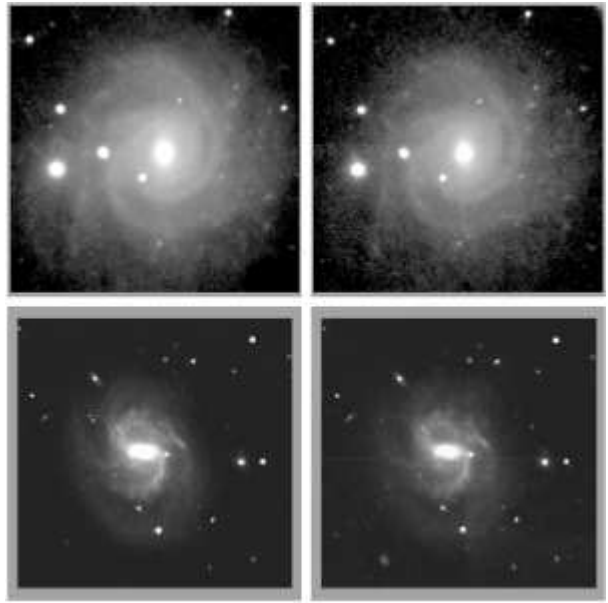,width=8cm}}
\end{center}
\caption{Surface brightness images for NGC 3344, $J$ and 
$K_{\rm s}$ (top panels, left to
right) and NGC 3686, $J$ and $K_{\rm s}$ (bottom panels, left to right).}
\label{Fig:sf3344-3686}
\end{figure}

\begin{figure}[!]
\begin{center}
\mbox{\epsfig{file=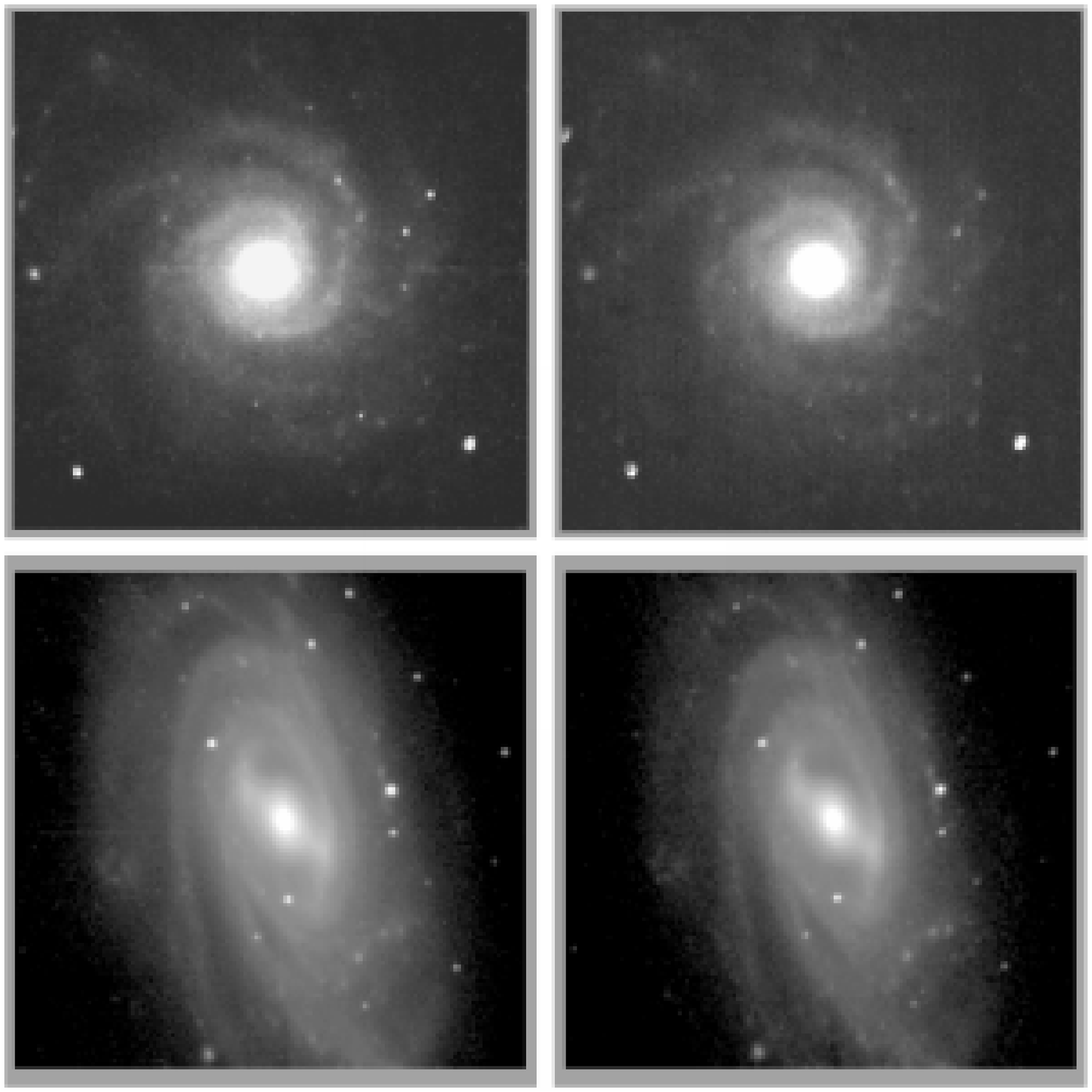,width=8cm}}
\end{center}
\caption{Surface brightness images for NGC 3938, $J$ and $K_{\rm s}$ (top panels, left to
right) and NGC 3953, $J$ and $K_{\rm s}$ (bottom panels, left to right).}
\label{Fig:sf3938-3953}
\end{figure}

\begin{figure}[!]
\begin{center}
\mbox{\epsfig{file=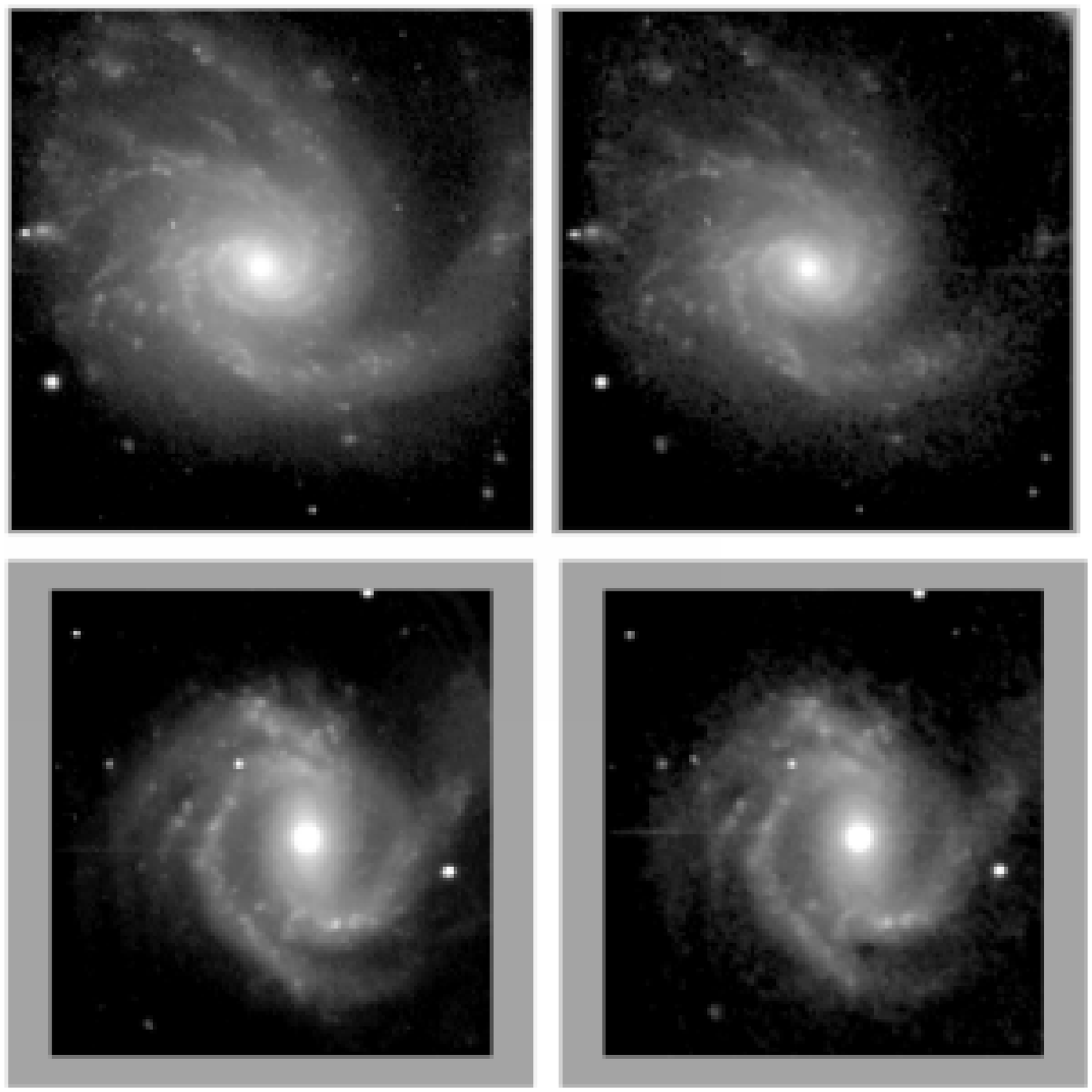,width=8cm}}
\end{center}
\caption{Surface brightness images for NGC 4254, $J$ and $K_{\rm s}$ (top panels, left to
right) and NGC 4303, $J$ and $K_{\rm s}$ (bottom panels, left to right).}
\label{Fig:sf4254-4303}
\end{figure}

\begin{figure}[!]
\begin{center}
\mbox{\epsfig{file=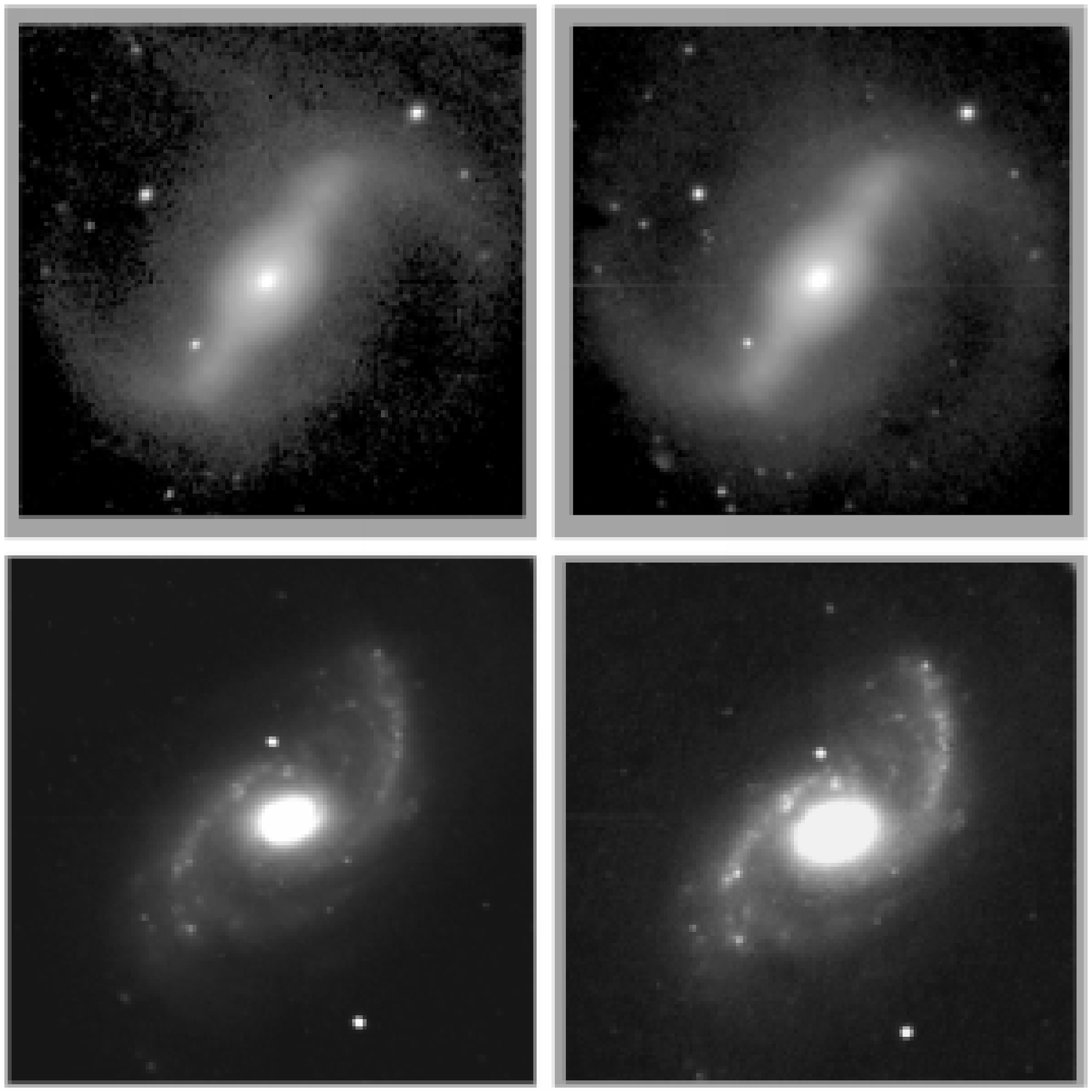,width=8cm}}
\end{center}
\caption{Surface brightness images for NGC 4314, $J$ and $K_{\rm s}$ (top panels, left to
right) and NGC 5248, $J$ and $K_{\rm s}$ (bottom panels, left to right).}
\label{Fig:sf4314-5248}
\end{figure}

\begin{figure}[!]
\begin{center}
\mbox{\epsfig{file=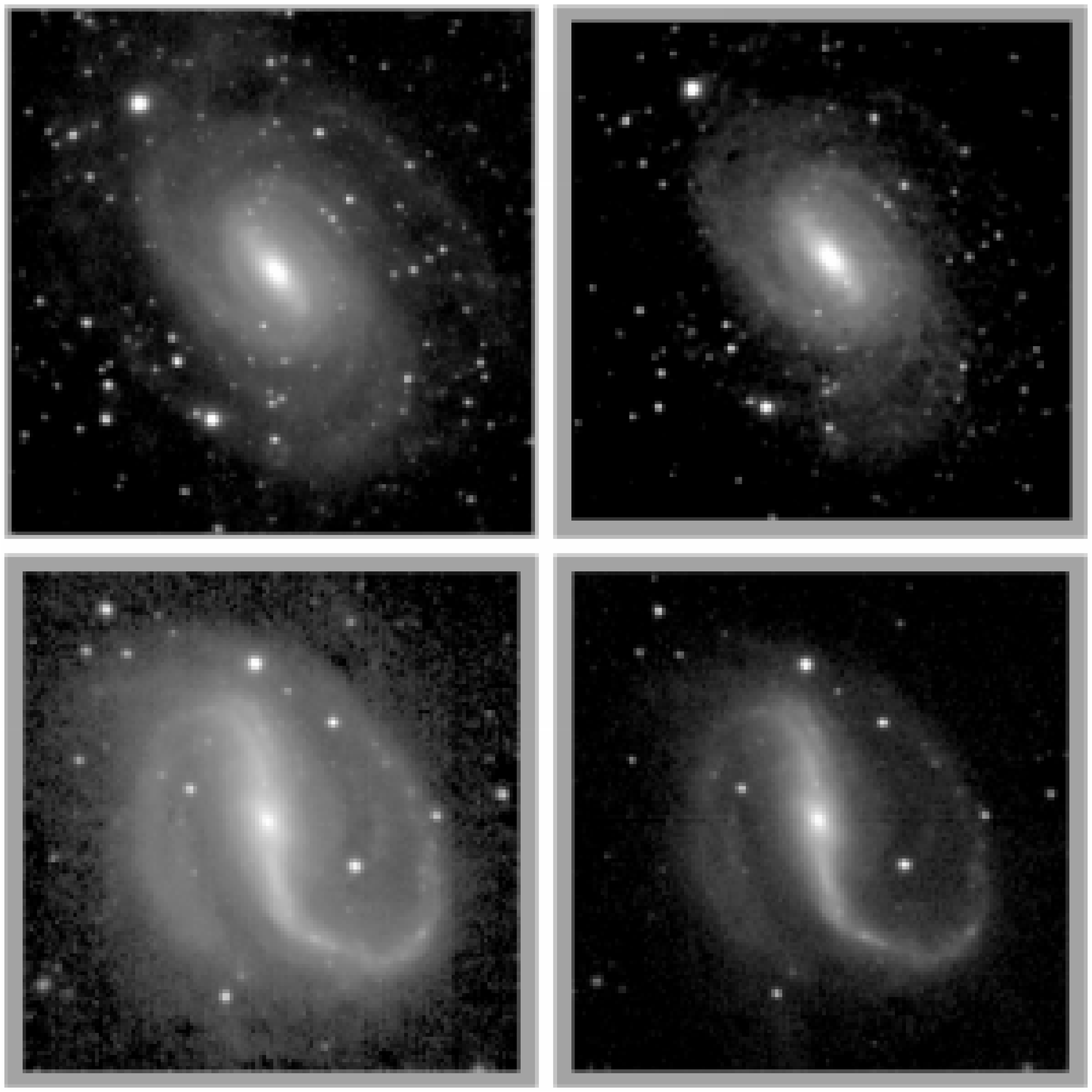,width=8cm}}
\end{center}
\caption{Surface brightness images for NGC 6384, $J$ and $K$ (top panels, left to
right) and NGC 7479, $J$ and $K_{\rm s}$ (bottom panels, left to right).}
\label{Fig:sf6384-7479}
\end{figure}

The observing strategy has been in all cases the classical ON--OFF method, in
which alternative exposures of the galaxy  and the adjacent sky are taken
until completing the total exposure time. The integration time of a single frame was determined by 
the linear part
of the well depth of the NICMOS detector and the sky flux, which is by far the
dominant contribution to the measured signal. Typical integration times per
frame range from 6 to 12 s in $K_{\rm s}$ and from 20 to 30 s in $J$. Several frames are
taken per pointing position until the grand total equals the timescale of the
variation of the sky background, which was set at two minutes with a sufficient
safety margin. This is then the time between nodding
positions (table \ref{Tab:caracteristicas}). The final
images showing the surface photometry of each galaxy are displayed in Figs.
$\ref{Fig:sf3344-3686}$, $\ref{Fig:sf3938-3953}$, $\ref{Fig:sf4254-4303}$,
$\ref{Fig:sf4314-5248}$ and $\ref{Fig:sf6384-7479}$.

\section{Data reduction and calibration}

The data have been reduced using several combinations of IRAF tasks. The
read-out mode has  in all cases been the standard Fowler mode, in which N
non-destructive reads are taken after the detector reset and  then integrated for
the required time and the detector  read $N$ times again. By so doing the
determination of the pedestal level of the signal is given by the average of
the first $N$ reads, and the image signal is defined by the difference between
the average of the last $N$ reads and that of the first ones. The S/N ratio of
the image is improved by this method .

We have followed the standard process for treating the data to correct
for bias and dark current, and flatfield calibration. Then we subtracted the
sky from each individual frame using a coadded image of the sky measurements
taken immediately before and after. Hence, we can correct for trends observed in
the background. Finally, the corrected object frames are aligned and averaged
to give the final product per galaxy and filter in adu. These are transformed
to flux-calibrated images using observations of several NIR standard
stars selected
from the UKIRT list (Hunt et al. 1998) observed during the night. 

In these calibrated coadded images we can estimate the sensitivity limits
defined as the signal  equivalent to 3 or 5 times the standard deviation. 
The results are given in table \ref{Tab:limite}.

\		 

\section{Components Decomposition}
\label{descomp}

One of our main purposes is to investigate the
presence and geometrical parameters of the principal structural components that
contribute to the observed surface brightness. Radial profiles
are the appropriate tool for this purpose and can be easily constructed with the
technique of ellipse fitting. This is done by means of the IRAF task ELLIPSE,
which computes the isophotal contours of the surface brightness images and fits
them to ellipses. ELLIPSE uses an algorithm developed by Jedrzejewski (1987).
As a result, one gets radial profiles of brightness, ellipticity and position
angles along the major axis of the ellipse family. This is formally equivalent
to first correcting for the inclination angle by deprojection of the galaxy
image and then getting those radial variations averaged over the azimuthal angle.
We have limited the scope of these profiles to the extent where the brightness
of the galaxies equals the sensitivity figures given in table \ref{Tab:limite}.
The profiles, then, will provide information about the morphological structure
of the galaxies.

Outer isophotes were used for determination of  global galactic parameters
such as ellipticity, position angle (hereafter, PA) and inclination angle, this
latter derived from the ellipticity. Galaxies with a lower inclination angle
have higher errors  because of the lack of a privileged
direction.

One can easily infer
the presence of the different structural components, discussed in the next
section, from the variation in these parameters with  galactic radius
(Varela et al. 1996; Wozniak et al. 1995). 

In table \ref{Tab:parametros} we have the results of the ELLIPSE task. The
ellipticity is defined as $b/a$, where $a $ and $b$
 are the major and minor axes of
the ellipses, respectively. 
The PA is the angle between the major axis of the ellipses and
north--south axis in the sky, measured from north to east. The values of each 
galaxy in both filters are in reasonably good
agreement,  their deviations
being  due mainly to errors in the ELLIPSE process,
possibly enhanced by the slightly different
sensitivities of the images in the two filters.

\begin {table*}[!]
\begin{center}
\begin {tabular}{cccccccccc}\hline

Galaxy & Filter & $R$($3\sigma$) & Ellipticity  & PA  & Inclination & Bar\ Ellipt. & Bar\ PA & m\_int &  m$_{2M}$ \\ 
       &        &   [arcsec] &         & [deg]&[deg]&  &[deg] \\
\hline
NGC 3344 & $J$   & 120& 0.09\( \pm  \)0.02 & 34.81\( \pm  \)4.48 & 24.79\( \pm  \)1.31 & 0.33\( \pm  \)0.01 & 84.73\( \pm  \)1.87 & 8.73 &  8.614   \\
	 & $K_{\rm s}$  & 101& 0.04\( \pm  \)0.11 & 35.60\( \pm  \)7.24 & 18.08\( \pm  \)6.45 & 0.26\( \pm  \)0.02 & 83.53\( \pm  \)2.81 & 8.17 &  7.755   \\
NGC 3686 & $J$   & 83 & 0.24\( \pm  \)0.03 & 21.64\( \pm  \)0.56 & 41.15\( \pm  \)0.07 & 0.59\( \pm  \)0.03 & 86.26\( \pm  \)3.25 & 9.18 &  9.474   \\
	 & $K_{\rm s}$  & 85 & 0.21\( \pm  \)0.01 & 22.08\( \pm  \)1.96 & 38.56\( \pm  \)0.77 & 0.61\( \pm  \)0.01 & 85.79\( \pm  \)1.30 & 8.44 &  8.504   \\
NGC 3938 & $J$   & 90 & 0.11\( \pm  \)0.01 & 65.95\( \pm  \)8.06 & 14.47\( \pm  \)0.58 & ---  & ---  			     & 9.24    &  ---	  \\
	 & $K_{\rm s}$  & 90 & 0.13\( \pm  \)0.00 & 62.45\( \pm  \)2.49 & 20.34\( \pm  \)0.25 & ---  & ---  			     & 8.06    &  ---	  \\
NGC 3953 & $J$   & 120& 0.50\( \pm  \)0.05 & 12.97\( \pm  \)3.91 & 60.18\( \pm  \)2.87 & 0.57\( \pm  \)0.03 & 55.90\( \pm  \)3.18 & 8.48 &  ---	  \\
	 & $K_{\rm s}$  & 120& 0.50\( \pm  \)0.07 & 12.31\( \pm  \)5.82 & 60.14\( \pm  \)4.31 & 0.58\( \pm  \)0.01 & 56.79\( \pm  \)1.53 & 7.69 &  ---	  \\
NGC 4254 & $J $  & 115& 0.27\( \pm  \)0.07 & 46.99\( \pm  \)9.59 & 43.21\( \pm  \)4.44 & ---  & ---  			     & 8.26    &  8.267   \\
	 & $K_{\rm s}$  & 115& 0.23\( \pm  \)0.08 & 39.41\( \pm  \)1.94 & 40.09\( \pm  \)0.48 & ---  & ---  			     & 7.74    &  7.308   \\
NGC 4303 & $J$   & 100& 0.16\( \pm  \)0.06 & 73.43\( \pm  \)4.81 & 33.69\( \pm  \)3.78 & 0.46\( \pm  \)0.00 &  2.71\( \pm  \)1.59 & 8.24 &  ---	  \\
	 & $K_{\rm s}$  & 110& 0.17\( \pm  \)0.04 & 78.48\( \pm  \)3.56 & 34.02\( \pm  \)2.36 & 0.27\( \pm  \)0.00 &  5.00\( \pm  \)3.58 & 7.57 &  ---	  \\
NGC 4314 & $J$   & 95 & 0.45\( \pm  \)0.08 & 33.68\( \pm  \)2.10 & 56.80\( \pm  \)4.91 & 0.66\( \pm  \)0.00 & 30.94\( \pm  \)1.01 & 9.44 &  8.768   \\
	 & $K_{\rm s}$  & 100& 0.41\( \pm  \)0.07 & 35.41\( \pm  \)5.82 & 54.29\( \pm  \)4.33 & 0.67\( \pm  \)0.09 & 35.51\( \pm  \)1.92 & 7.48 &  7.811   \\
NGC 5248 & $J $  & 110& 0.40\( \pm  \)0.00 & 47.17\( \pm  \)1.43 & 53.35\( \pm  \)0.31 & ---  & ---  			     & 8.24    &  ---	  \\
	 & $K_{\rm s}$  & 105& 0.38\( \pm  \)0.02 & 48.51\( \pm  \)4.25 & 51.87\( \pm  \)1.28 & ---  & ---  			     & 7.44    &  ---	  \\
NGC 6384 & $J$   & 115& 0.48\( \pm  \)0.03 & 22.34\( \pm  \)2.14 & 58.88\( \pm  \)2.12 & 0.38\( \pm  \)0.07 & 35.34\( \pm  \)3.56 & 9.02 &  ---	  \\
	 & $K$   & 115& 0.31\( \pm  \)0.06 & ---                 & 47.08\( \pm  \)3.68 & 0.40\( \pm  \)0.09 & 29.77\( \pm  \)2.81 & 9.22 &  --      \\
NGC 7479 & $J$   & 110& 0.25\( \pm  \)0.03 & 37.99\( \pm  \)5.30 & 41.67\( \pm  \)2.26 & 0.68\( \pm  \)0.01 &  1.52\( \pm  \)1.10 & 8.81 &  9.341   \\
	 & $K_{\rm s}$  & 110& 0.24\( \pm  \)0.03 & 35.13\( \pm  \)9.60 & 41.26\( \pm  \)2.86 & 0.70\( \pm  \)0.29 &  0.43\( \pm  \)2.81 & 8.16 &  8.360   \\

\hline  
\end {tabular} 
\end{center}
\caption{Parameters derived from the ellipse fitting and the integrated
magnitude over an aperture of radius $R$($3\sigma$). For comparison, 2MASS
integrated magnitudes have been included in the last column.}

\label{Tab:parametros}

\end {table*}

We can compare our results for the inclination angle and PA of each galaxy listed
in table \ref{Tab:parametros} with the values quoted in RC3. The agreement is generally
good. NGC 4314 exhibits the largest differences certainly
because of its large size
in relation with the instrument FOV. We will not extend further the comparison
 between the NIR images presented here and the optical images in the
 RC3, since the component descomposition is not performed in the optical images.
 The reasons for this are to be found in the non-uniformity of the optical images
 and the expected spurious effects due to observational circunstances beyond our 
 control. However, it is clear that
 the NIR images more closely represent the true stellar distribution caused by the smaller
 value of the extintion. It is then expected that the geometrical features of the
 main structural components will be somewhat different in the two spectral
 regimes.

Using the results of the ellipse fitting, we have measured the integrated
magnitudes of the sample in both filters (see table \ref{Tab:parametros})
within a circular aperture of the radius at which the surface brightness from
the object equals that of the sky at $3\sigma$ over its mean value (see table
\ref{Tab:limite}). These values are in fairly good agreement with those
from the 2MASS data (Skrutskie et al. 1995), taken from NED (see table
\ref{Tab:parametros}). However, our images extend in general to  larger radii
than the fixed scale of 80 arcsec quoted in NED.

In those galaxies where a clear indication exists of the presence of a large
central bar, either from the images themselves and/or from the sudden change in
the radial profiles. We used an additional method to estimate the geometrical
parameters of the bar itself. The global PA and inclination angle of the object galaxy are used to
deproject the galaxy image in order to measure quantities in the proper
galactic plane.. Columns  6 and 7 of table
\ref{Tab:parametros} show the values of ellipticity and PA for the bar taken from
the ELLIPSE results, after deprojection, and measured at the galactic radius
spanned by the bar.

Once the radial profiles have been obtained and the general parameters
calculated, we can now proceed with the decomposition of the averaged
brightness profile into different structural components, each with a
separate contribution to the observed flux. There are a large number of
similar researches in the literature where structural decomposition of external
galaxies are attempted in various ways (Simien \& Michard 1990; Prieto et al.
1990, 1992; Byun \& Freeman 1995), but most of them are based on visible data
that do not so closely resemble the true stellar distribution because of
their higher value of the extinction. Previous NIR decomposition can be found in
de Jong (1996), Peletier \& Balcells (1997), Moriondo et al. (1998), Seigar \&
James (1998) and references therein. For the different component
identification we have followed the technical approach of Prieto et al. (2001)
where the changes in the radial profiles of ellipticity, PA and B4 are used to
identified the radial interval over which the given component extends. 

Then, the analytical functions, described in the
table \ref{Tab:parametros1} and in Aguerri (1998), are fitted to the brightness 
radial profile within that interval. This technique would be of little use
in the case of strong bars since these structures only occur over a narrow
range of azimuthal angles. Hence, the use of azimuthally averaged profiles, as
is the case for the results of the ellipse fitting, could well  mask the bar
against the averaged background. We will come back to these cases at the end of
this section.

\begin {table*}[!]
\begin{center}
\begin {tabular}{ccc}\hline
Component& equation&Reference \\ \hline
 Bulge&

 $I_{b} (r) = I_{0b} 10^{-b_n[(r/r_e)^{1/n} - 1]}$,

$ b_n=0.868n-0.142$,
& Sersic 1968; Aguerri 1998\\ 
Disco&

$I_d (r) = I_{0d} \exp\left(-r/h\right)$,

& Freeman 1970\\ 
Elliptical Bar&

 $I_{ba} (x,y) = I_{0,ba} \sqrt{1-\left( \frac{x}{a_{bar}}\right) ^2 -
\left(\frac{y}{b_{bar}}\right)^2}$,

& Prieto et al. 1997, 2001; Freeman 1966\\
Flat Bar&

 $I_{ba} (r) = \frac{I_{0,ba}}{1 + \exp\left(\frac{r-\alpha}{\beta}\right)}$,

& Prieto et al. 2001\\ 
Lens&

 $I_l (r) = I_{0l} \left[1- \left(\frac{r}{r_l}\right)^2\right]$,

& Duval \& Athanasoula 1983\\ 

Ring&

$I_r (r) =I_{0r} \exp\left(-\frac{(r - r_{ro})^2}{2\sigma^2}\right)$,

& Buta 1996\\ \hline

\hline  
\end {tabular} 
\end{center}
\caption{Parameters derived from the ellipse fitting and the integrated
magnitude over an aperture of radius $R$($3\sigma$). For comparison, 2MASS
integrated magnitudes have been included in the last column.}

\label{Tab:parametros1}

\end {table*}

Our procedure to fit the several morphological components to the measured
brightness profile is similar to that described in Prieto et al. (2001). Once
the components have been identified in the radial profiles of several
parameters coming from the ellipse fitting, we first fix the disc in the outer
parts of the galaxy, where contamination from other components can be
neglected. Next, this disc model is subtracted from the original profile and
we fit the bulge to the residuals with the best Sersic law ($n=1$--4). These
operations are iterated until convergence, defined as the difference between two
consecutive set of parameters being less than the measured noise,
is reached.

After the  disc--bulge pair has been set up, we proceed with the secondary
components, by fitting the residuals of the subtraction of the disc--bulge
model from the brightness profile to the spatial extent indicated in the radial
profiles of the B4, PA and ellipticity parameters. In galaxies with prominents
bars, such as NGC 7479 or NGC 4314, we have tried  two different radial
profiles for the fitting. In addition of using the azimuthally averaged
brightness profiles resulting from the ellipse fitting, we have also built
another two averaged brightness profile by computing the radial profiles by
averaging in a narrower range of azimuthal angles: either profiles along the
bar direction or perpendicular to the bar. A  disc--bulge 
pair is now fitted to
the profiles averaged along the minor axis of the bar where the bar contribution
is low. Then, this model is subtracted from the radial profile averaged along the
major axis of the bar direction and the residuals matched with a bar model. As
before, this procedure is iterated until convergence, defined as in the previous
case, is reached. 
In the following sections we will refer to the decomposition results from the
ELLIPSE fitting as {\it isophotal parameters} and as {\it fitting parameters}
to those of this last technique. In the case of the fitting parameters for
the bar, we can also fit a bar profile  along both the major and minor axes of
the bar once the  disc--bulge pair is fixed.


\section{Results}

The numerical results of our component decomposition are summarized in tables \ref{Tab:parametros2}
and \ref{Tab:parametros3}. Historically, the
standard function for approximating galactic bulges has been the $r^{1/4}$
law, but
it has been suggested that some bulges might be better represented by an
$r^{1/n}$ law (Caon et al. 1993; Andredakis et al. 1995). When $n$ increases, the
luminosity profiles are more centrally concentrated. For this reason,
 early-type galaxies show larger values of $n$ than late
 types (Andredakis et al. 1995).
In this respect, observational effects can affect the results of the fit.
For instance, the {\it seeing} size, if comparable with the galaxy diameter,
can distort the concentration of the surface brightness (Trujillo et al. 2001).
This should not be the case in our sample since all the galaxy bulges are much
bigger than the seeing. In our fits, the best results for the whole sample is
obtained with a value of $n$ equal to 1, but we cannot correlate $n$ with the
morphological type because we have only Sc or Sbc galaxies (late-type
galaxies) in this first sample of objects.


\section{Conclusion}
 
Most of the galaxies in the sample are of late type (bc or c), 
NGC 4314 being the only example of an early-type (SBa) object. Hence, we cannot infer
any correlations between morphological type and the component parameters. This will be done in 
subsequent works with a larger sample of
morphological types.  The majority of the
galaxies in the sample have $T$ in the range  4--5,
see table \ref{Tab:rc3}. Only NGC 4314 and NGC 6384
are earlier galaxies with  $T$ 1.0 and 3.6
respectively.

The  bulge/disc ratio ($B/D$), defined in  equation \ref{eq:ec12} (Mihos et al.
1994), 

\begin{equation}
 B/D= \frac{1}{0.28}\frac{r_e}{h}\frac{\Sigma_e}{\Sigma_o},
\label{eq:ec12}
\end{equation}

is rather low (see table \ref{Tab:razon}), showing that the inner parts of the galaxy are
dominated by components other than the central bulge,
most probably the bar and ring.

It is worth  noting, however, that NGC 4314 has the highest B/D ratio, with a value 
around of 0.6, indicating its early type. This galaxy
has a spike 
in the ellipticity and PA profiles around 10$''$. This peak points to the 
presence of an additional bar in the inner parts of the galaxy (Aguerri 1998). 
The existence of this secondary bar in NGC 4314 has been also proposed by Erwin 
\& Sparke (2002) analysing images from
the Hubble Space telescope (see their fig.  2b). The secondary bar has been 
fitted to the residuals after removal of the main components in the averaged 
brightness radial profile.

\begin{table}
\begin{center}
\begin{tabular}{ccc} \hline
Galaxy & $J$ & $K_{\rm s}$ \\ \hline \hline
NGC 3344 & 0.12 &  0.10   \\
NGC 3686 & 0.11 &  0.09  \\ 
NGC 3938 & 0.14 &  0.15  \\ 
NGC 3953 & 0.17 &  0.14  \\ 
NGC 4254 & 0.38 &  0.16  \\ 
NGC 4303 & 0.12 &  0.19  \\ 
NGC 4314 & 0.64 &  0.59  \\ 
NGC 5248 & 0.33 &  0.42  \\ 
NGC 6384 & 0.31 &  0.34  \\ 
NGC 7974 & 0.21 &  0.17  \\ 
\hline
\end{tabular} 
\end{center}
\caption{Bulge/disc ratio, as defined in  equation \ref{eq:ec12}}
\label{Tab:razon}
\end{table}

The resulting inclinations and PAs are in agreement with the data of other
authors, in particular with the RC3. The major differences are between those galaxies with low PAs, in which the
measurements are more difficult because of the poorer definition of this
parameter. This is equally true for the inclination angles. There are minor
discrepancies between the numbers in each filter, with some departures in NGC
3938 because of its shorter exposure time in $K_{\rm s}$. On the other hand, NGC 4303
and NGC 4314 show the greatest difference in inclination compared with the RC3
values in both filters. NGC 4303 has outer arms that are
fainter in the NIR than in the
optical bands, so we might be underestimating the outer part of the galaxy (on
the other hand, we are comparing visible and NIR data). The same reason applies
to NGC 4314. In general, all the galaxy images are bigger in the optical band,
the images being extracted from the DSS, than in this work. But the particular
orientation of the outer arms and the central bar causes these larger
departures in these two galaxies.
Some galaxies present a ring in RC3 (NGC 3686,
NGC3953, NGC 5248 and NGC 6384), in NIR their are
very faint in the images and could be mixed with the
arms.

\begin{figure}[h]
\begin{center}
\mbox{\epsfig{file=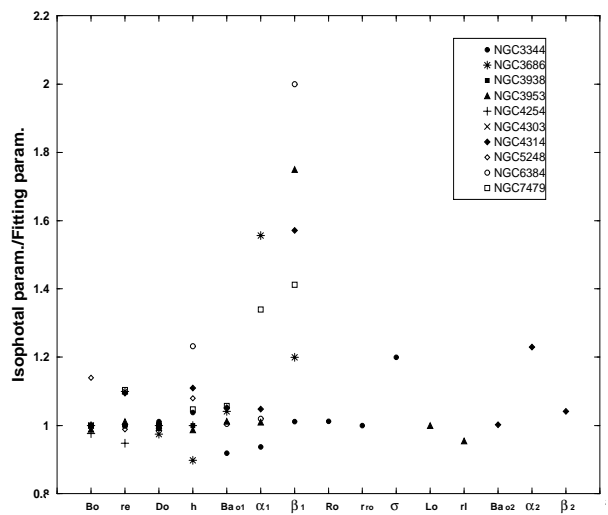,width=8.cm}}
\end{center}
\caption{Correlation between the two sets of parameters for the brightness
profiles: isophotal parameters (i) vs.~fitting parameters (f). The graphic represents the 
ratio between the same parameters of each galaxy's surface brightness decomposition
calculated using the two methods. See caption of table 5
for an explanation of the parameter names in the $x$-axis.}
\label{Fig:fitting}
                                
\end{figure}

The two methods that we have used to fit the parameters of the brightness
distribution for the morphological components, namely the {\it isophotal
parameters} and the {\it fitting parameters}, give very similar results, as can
be seen in Fig. \ref{Fig:fitting}. This can be attributed to
the small influence of
the bar in the overall brightness distribution outside  the very central
regions. In addition, part of the observed differences are certainly due to
innaccuracies inherent in the fitting procedures. We can see that galaxies with
strong bars (such as NGC 7479 and NGC 4314) have their isophotal
bar parameters overestimated. This happens because of distortions in the ellipse
fitting due to the presence of the spiral arms in the interarm region near 
the bar (NGC 6384, NGC 3953 and NGC 3686; see Figs.  \ref{Fig:sf3344-3686},
\ref{Fig:sf3938-3953} and \ref{Fig:sf6384-7479}). 

We have obtained the best results with $n=1$ in the Sersic law for all the
bulge fittings. Prieto et al. (2001) have argued that $n$ increases with the
wavelength of the  filters. We cannot discuss this effect since our
filters are rather close each other in wavelength. Only NGC 4314 (the 
early-type galaxy) presents a possible $n=2-3$ class bulge, but in the ellipticity
profile there is a peak around $r=10''$, which
indicates a secondary bar (Erwin \& Sparke 2002; Aguerri 1998). After the
removal of this component, the profile of the bulge is then back of type $n=1$.
The uniqueness of this result may be questioned, since we would have
 obtained a similarly good agreement with the observed profile using a
bigger and triaxial bulge instead of a secondary bar. However Erwin \& Sparke
(2002), using deep images taken with the Hubble Space Telescope, have also found these
inner structures. Some authors (Andredakis et al. 1995; Moriondo et al. 1998)
have reported a correlation between the exponent n in the Sersic law with the
morphological type of the galaxy, which we cannot check here
because of the
lack of coverage of this parameter.

It is noteworthy that NGC 4314 and NGC 7479 present the strongest bars in the
sample, and that both have a LINER in their nuclei. This fact is discussed by Martin
(1995), Athanasoula (1982) and Friedli \& Berz (1993). Finally, all the bars
found in this work are best fitted assuming that they are flat-type bars
that follow the equation in table \ref{Tab:parametros1}.

\begin{acknowledgements}

We are thankful to Drs J.A. L\'opez Aguerri and A. Cabrera for many helpful and
useful conversations during the preparation of this manuscript. To Norberto
Castro-Rodr\'\i guez for his helpful in the fitting of the profiles. The CST is
operated on the island of Tenerife by the Instituto de Astrof\'{\i}sica de
Canarias at the Spanish Observatorio del Teide of the Instituto de
Astrof\'{\i}sica de Canarias. This article makes use of data products from 2MASS, which is
a joint project of the Univ. of Massachusetts and the Infrared Processing
and Analysis Center, funded by the NASA and the NSF. This work has been partially funded by the
Spanish {\it Plan Nacional de Astronom\'{\i}a y Astrof\'{\i}sica}, project
AYA2000-2046-C02-02 and the {\it Universidad de La Laguna-Cajacanarias}
fellowship (Canary Islands, Spain).

\end{acknowledgements}

\begin{table*}

\begin{center}

\begin{tabular}{cccccccc}\hline

Object & \multicolumn{2}{c}{Bulge} & \multicolumn{2}{c}{Disc} & \multicolumn{3}{c}{Bar}  \\
& $-2.5\log I_{0b}$ & $r_e$ & $-2.5\log I_{0d}$ & $h$ & $-2.5\log I_{0,ba}$ & $\alpha$ & $\beta$  \\
& mag  \ [$B_{\rm o}$]& arcsec & mag  \ [$D_{\rm o}$]& arcsec & mag  \ [Ba$_{o1}$]& arcsec & arcsec  \\ 
\hline 
3344 $J$ & 16.80$\pm$0.06 & 3.80$\pm$1.20 & 18.35$\pm$0.01 & 41.77$\pm$1.0  & 18.30$\pm$0.10 & 12.00$\pm$1.0 & 4.00$\pm$0.5   \\
$K_{\rm s}$     & 15.84$\pm$0.06 & 3.50$\pm$1.45 & 17.48$\pm$0.01 & 43.79$\pm$3.0  & 17.30$\pm$0.08 & 13.00$\pm$1.0 & 4.00$\pm$0.5  \\ 
3686 $J$ & 18.38$\pm$0.08 & 3.62$\pm$0.45 & 19.26$\pm$0.09 & 30.94$\pm$5.0  & 19.05$\pm$0.03 & 16.00$\pm$0.5 & 5.00$\pm$0.5 \\
$K_{\rm s}$     & 17.77$\pm$0.10 & 3.59$\pm$0.10 & 18.35$\pm$0.20 & 28.43$\pm$9.0  & 18.5 $\pm$0.05 & 21.00$\pm$0.5 & 5.00$\pm$0.5 \\
3938 $J$ & 18.38$\pm$0.01 & 6.30$\pm$0.30 & 18.24$\pm$0.05 & 29.70$\pm$0.2  \\
$K_{\rm s}$     & 17.40$\pm$0.05 & 6.83$\pm$0.20 & 17.35$\pm$0.05 & 31.85$\pm$0.3  \\
3953 $J$ & 17.81$\pm$0.20 & 6.27$\pm$0.09 & 19.60$\pm$0.09 & 64.00$\pm$1.5  & 19.60$\pm$0.06 & 44.00$\pm$0.5 & 4.00$\pm$0.5  \\
$K_{\rm s}$     & 16.97$\pm$0.20 & 5.87$\pm$0.15 & 18.70$\pm$0.06 & 64.64$\pm$3.0  & 18.45$\pm$0.08 & 45.00$\pm$0.5 & 4.00$\pm$0.5  \\
4254 $J$ & 17.83$\pm$0.15 & 9.56$\pm$1.00 & 17.93$\pm$0.10 & 30.33$\pm$1.6  &                & 		   &               \\
$K_{\rm s}$     & 16.72$\pm$0.20 & 7.01$\pm$2.00 & 17.47$\pm$0.02 & 46.41$\pm$0.3  & 	           & 		   & 		   \\
4303 $J$ & 16.02$\pm$0.04 & 3.86$\pm$0.70 & 18.67$\pm$0.05 & 69.61$\pm$1.0  & 18.05$\pm$0.80 & 26.00$\pm$0.5 & 9.00$\pm$1.0 \\
$K_{\rm s}$     & 15.02$\pm$0.06 & 3.49$\pm$0.10 & 17.83$\pm$0.06 & 55.10$\pm$0.5  & 17.55$\pm$0.07 & 22.00$\pm$0.5 & 5.00$\pm$0.5 \\
4314 $J$ & 18.01$\pm$0.07 & 6.02$\pm$0.70 & 20.52$\pm$0.08 & 45.06$\pm$1.0  & 19.95$\pm$0.10 & 73.00$\pm$1.0 & 7.00$\pm$0.5  \\
$K_{\rm s}$     & 15.82$\pm$0.20 & 5.48$\pm$0.10 & 18.35$\pm$0.10 & 42.92$\pm$2.5  & 17.85$\pm$0.10 & 76.60$\pm$1.0 & 7.00$\pm$0.5  \\
5248 $J$ & 17.55$\pm$0.04 & 7.15$\pm$0.10 & 18.05$\pm$0.05 & 29.19$\pm$1.8  \\
$K_{\rm s}$     & 16.71$\pm$0.04 & 8.38$\pm$0.15 & 17.40$\pm$0.10 & 33.21$\pm$3.0  \\
6384 $J$ & 17.78$\pm$0.08 & 6.22$\pm$1.00 & 19.35$\pm$0.30 & 43.10$\pm$0.9  & 19.20$\pm$0.05 & 25.50$\pm$0.5 & 2.50$\pm$1.0 \\
$K$      & 16.98$\pm$0.30 & 6.06$\pm$1.50 & 19.36$\pm$0.10 & 58.00$\pm$10.0 & 17.95$\pm$0.06 & 26.00$\pm$0.5 & 4.50$\pm$3.0 \\
7479 $J$ & 18.20$\pm$0.02 & 6.02$\pm$0.20 & 20.16$\pm$0.15 & 60.30$\pm$3.0  & 18.55$\pm$0.05 & 50.00$\pm$0.5 & 7.50$\pm$0.5 \\
$K_{\rm s}$     & 16.7 $\pm$0.04 & 4.80$\pm$0.35 & 18.90$\pm$0.15 & 60.13$\pm$0.5  & 17.40$\pm$0.05 & 47.00$\pm$1.0 & 8.50$\pm$0.5 \\
\hline
\end{tabular} 
\caption{Isophotal parameters of each component per individual galaxy and
filter. The root {\it NGC} has been removed from the galaxy
name for simplicity. See text for description. The names in brackets in
the header section are those used in fig. \ref{Fig:fitting} for that
parameter.}

\label{Tab:parametros2}
\end{center}

\end{table*}

\begin{table*}
\footnotesize{
\begin{narrow}{-0.2cm}{-0.5cm}

\begin{center}

\begin{tabular}{ccccccccc}\hline

Object &  \multicolumn{3}{c}{Ring} & \multicolumn{2}{c}{Lens} & \multicolumn{3}{c}{Secondary Bar} \\
&  $-2.5\log I_{0r}$ & $r_{ro}$ & $\sigma$ & $-2.5\log I_{0l}$ & $r_l$ & $-2.5\log I_{0,ba}$ & $\alpha$ & $\beta$ \\
& mag  \ [$R_{\rm o}$]& arcsec & arcsec & mag  \ [$L_{\rm o}$]& arcsec & mag  \ [Ba$_{o2}$]& arcsec & arcsec \\ 
\hline 
3344 $J$ &  20.75$\pm$0.2 & 26.50$\pm$0.5 & 7.00$\pm$1.0  \\
$K_{\rm s}$     &  19.50$\pm$0.1 & 25.00$\pm$0.5 & 5.00$\pm$2.0  \\ 
3953 $J$ &                &               & & 19.50$\pm$0.2 & 22.00$\pm$0.5 \\
$K_{\rm s}$     &                & 	          & & 18.60$\pm$0.2 & 24.00$\pm$1.0 \\
4254 $J$ &                 & 		   &              & 19.80$\pm$0.2 & 16.00$\pm$0.1 & 7.00$\pm$0.5 \\
$K_{\rm s}$     &  	           & 		   & 		  & 18.35$\pm$0.2 & 12.50$\pm$0.7 & 5.00$\pm$1.0 \\
4314 $J$ &                &               & & &  & 19.50$\pm$0.05 & 23.50$\pm$0.1 & 5.00$\pm$0.5 \\
$K_{\rm s}$     &               &               & & &  & 17.02$\pm$0.07 & 19.50$\pm$2.0 & 6.00$\pm$0.5 \\
\hline
\end{tabular} 
\caption{Isophotal parameters of each component per individual galaxy and
filter. The root {\it NGC} has been removed from the galaxy
name for simplicity. See text for description. The names in brackets in
the header section are those used in fig. \ref{Fig:fitting} for that
parameter. The galaxies of the sample without a componet are not presented in this table.}

\label{Tab:parametros3}
\end{center}
\end{narrow} 
}
\end{table*}


\begin{thebibliography}{99}


\bibitem [\protect\citeauthoryear{}{}]{} Aguerri, J.
L. 1998, Thesis, Universidad de La Laguna (Spain).

\bibitem [\protect\citeauthoryear{}{}]{} Andredakis, Y. C., Peletier, R. F., \& Balcells, M.
1995, MNRAS, 275, 874

\bibitem [\protect\citeauthoryear{}{}]{} Athanasoula, E., Bosma, A., Creze, M., \& 
Schwarz, M. P. 1982, A\&A, 107, 101

\bibitem [\protect\citeauthoryear{}{}]{} Buta, R. 1996, ASP Conf. Ser. 91:
IAU Collq. 157: Barred galaxies, p.11

\bibitem [\protect\citeauthoryear{}{}]{} Byun, Y., \& Freeman, K. 1995, ApJ, 448, 563

\bibitem [\protect\citeauthoryear{}{}]{} Caon, N., Capaccioli, M., \& D'Onofrio,
M. 1993, MNRAS, 265, 1013

\bibitem [\protect\citeauthoryear{}{}]{} Duval, M. F., \& Athanasoula, E.
1983, A\&A, 121, 297

\bibitem [\protect\citeauthoryear{}{}]{} Erwin, P.,
\& Sparke, L. 2002, AJ, 124, 65

\bibitem [\protect\citeauthoryear{}{}]{} Freeman, K. C. 1966, MNRAS, 133, 47

\bibitem [\protect\citeauthoryear{}{}]{} Freeman, K. C. 1970, ApJ, 160, 811

\bibitem [\protect\citeauthoryear{}{}]{} Friedli, D., \& Berz, W. 1993, A\&A,
268,65

\bibitem [\protect\citeauthoryear{}{}]{} Giovanardi, C., \& Hunt, L. K. 1996, AJ, 111, 1086

\bibitem [\protect\citeauthoryear{}{}]{} Huchtmeier, W. K., \& Richter, O.
G. 1989, A general catalogue of HI observations of galaxies, Springer-Verlag,
New York.

\bibitem [\protect\citeauthoryear{}{}]{} Hunt, L. K., Mannuci, F., Testi, L.,
Milgliorini, S., \& Stanga, R. M. 1998, AJ, 115, 2594

\bibitem [\protect\citeauthoryear{}{}]{} Hunt, L. K., Malkan, M. A., Salvati, M., Mandolesi, N.,
Palazzi, E., \& Wade, R. 1997, ApJS., 108, 229

\bibitem [\protect\citeauthoryear{}{}]{} Jedrzejewski, R. J. 1987, IAUS, 127,
37 

\bibitem [\protect\citeauthoryear{}{}]{} De Jong, R. S. 1996, A\&A, 118, 557

\bibitem [\protect\citeauthoryear{}{}]{} De Jong, R. S. 1996, A\&A, 313, 45

\bibitem [\protect\citeauthoryear{}{}]{} De Jong, R. S. 1994, A\&A Sup., 106, 451

\bibitem [\protect\citeauthoryear{}{}]{} De Jong, R. S., \& van der Kruit, P. C. 1994, A\&AS, 106,
451

\bibitem [\protect\citeauthoryear{}{}]{} Kennicutt, R. C. 1989, ApJ, 344, 685

\bibitem [\protect\citeauthoryear{}{}]{} Kennicutt, R. C. 1983, ApJ, 272, 54

\bibitem [\protect\citeauthoryear{}{}]{} Martin, P. 1995, AJ, 109, 2428

\bibitem [\protect\citeauthoryear{}{}]{} Mihos, C., \& Hernquist, L. 1994, ApJ, 425, 13

\bibitem [\protect\citeauthoryear{}{}]{} Moriondo, G., Giovanardi, C., \& Hunt,
L. K. 1998, A\&AS, 130, 81

\bibitem [\protect\citeauthoryear{}{}]{} Peletier, R. F., \& Balcells, M. 1997, NewA, 1, 349

\bibitem [\protect\citeauthoryear{}{}]{} Peletier, R .F., \& Balcells, M. 1996, AJ, 111, 6

\bibitem [\protect\citeauthoryear{}{}]{} Peletier, R .F., Valentijn, E. A., Moorwood, A. F. M., \&
Freudling, W. 1994, A\&AS, 108, 621

\bibitem [\protect\citeauthoryear{}{}]{} Prieto, M., Aguerri, J.  L., Varela,
A. M., \& Mu\~noz--Tu\~n\'on, C. 2001, A\&A, 367, 405

\bibitem [\protect\citeauthoryear{}{}]{} Prieto, M., Gottesman, S. T., Aguerri,
J. L., \& Varela, A. 1997, AJ, 114, 1413

\bibitem [\protect\citeauthoryear{}{}]{} Prieto, M., Beckman, J. E., Cepa, J., \&  Varela, A. 1992,
A\&A, 257, 85 

\bibitem [\protect\citeauthoryear{}{}]{} Prieto, M., Cepa, J., Beckman, J. E., Varela, A., \& 
Mu\~noz--Tu\~n\'on, C.  1990, Ap\&SS, 170, 225 

\bibitem [\protect\citeauthoryear{}{}]{} Seigar, M. S., \& James, P. A. 1998, MNRAS, 299, 672

\bibitem [\protect\citeauthoryear{}{}]{} Sersic, J. L. 1968,  Atlas de galaxias australes C\'ordoba:Observatorio astron\'omico

\bibitem [\protect\citeauthoryear{}{}]{} Simien, F., \& Michard, R. 1990, A\&A, 227, 11

\bibitem [\protect\citeauthoryear{}{}]{} Skrutskie, M. F., et al. 1995, AAS, 187, 7507

\bibitem [\protect\citeauthoryear{}{}]{} Trujillo, I., Aguerri, J. L., Cepa,
J., \& Gutierrez, C. M. 2001, MNRAS, 328, 977

\bibitem [\protect\citeauthoryear{}{}]{} Valentijn, E. A.  1994, MNRAS, 266, 614

\bibitem [\protect\citeauthoryear{}{}]{} Van der Kruit, P. C., \& Searle, L.  1982, A\&A, 110,79

\bibitem [\protect\citeauthoryear{}{}]{} Van der Kruit, P. C., \& Searle, L.  1981, A\&A, 95, 116

\bibitem [\protect\citeauthoryear{}{}]{} Varela, A. M., Mu\~noz--Tu\~n\'on, C.,
\& Simmoneau, E.  1996, A\&A, 306, 381

\bibitem [\protect\citeauthoryear{}{}]{} de Vaucouleurs, G. 1991, RC3-"Third Reference Catalogue of Bright
Galaxies", Springer-Verlag 

\bibitem [\protect\citeauthoryear{}{}]{} Wozniak, H., Friedli, D., Martinet, L.,
\& Bratschi,  1995, A\&AS, 111, 115

\end{thebibliography}
\end{document}